\documentclass[conference]{IEEEtran}
\IEEEoverridecommandlockouts

\usepackage{amsmath,amsfonts}
\usepackage{algorithm}
\usepackage{algorithmicx}
\usepackage{array}
\usepackage{textcomp}
\usepackage{stfloats}
\usepackage{url}
\usepackage{verbatim}
\usepackage{graphicx}
\usepackage{cite}
\usepackage{amsmath}
\usepackage[acronym]{glossaries}
\usepackage{algpseudocode}
\usepackage{cite}
\usepackage{color, soul}
\usepackage[acronym]{glossaries}
\usepackage{lipsum}
\newacronym{SOTA}{SOTA}{state of the art}
\DeclareMathOperator*{\argmax}{arg\,max}
\DeclareMathOperator*{\argmin}{arg\,min}

\newacronym{AUD}{AUD}{active users detection}
\newacronym{MTC}{MTC}{machine type communications}
\newacronym{SNR}{SNR}{signal-to-noise-ratio}
\newacronym{mMTC}{mMTC}{massive \gls{MTC}}
\newacronym{BS}{BS}{base station}
\newacronym{MIMO}{MIMO}{multiple input multiple output}
\newacronym{CS}{CS}{compressed sensing}
\newacronym{OMP}{OMP}{orthogonal matching pursuit}
\newacronym{ROC}{ROC}{receiver operating characteristic}
\newacronym{FPR}{FPR}{false positive rate}
\newacronym{FNR}{FNR}{false negative rate}
\newacronym{CI}{CI}{coherence interval}
\newacronym{SIC}{SIC}{successive interference cancellation}
\newacronym{IoT}{IoT}{internet of things}
\newacronym{QoS}{QoS}{quality of service}
\newacronym{UR}{UR}{ultra-reliability}
\newacronym{non-iterative}{UMC-OMP}{Unconnected Multi-Channel \gls{OMP}}
\newacronym{iterative}{IMC-OMP}{Iterative Multi-Channel \gls{OMP}}
\newacronym{BA}{BA}{balanced inaccuracy}

\usepackage{graphicx}
\usepackage[colorlinks=true, allcolors=blue]{hyperref}
\usepackage{amsthm,amssymb}

\algnewcommand{\Initialize}[1]{%
  \Statex \hspace{-6mm}\textbf{Initialize:}
}

\begin{document}

\title{Assessment of the Sparsity-Diversity Trade-offs in Active Users Detection for mMTC with the Orthogonal Matching Pursuit}


 \author{
    \IEEEauthorblockN{Gabriel Martins de Jesus\IEEEauthorrefmark{1}, Onel Alcaraz L{\'opez}\IEEEauthorrefmark{1}, Richard Demo Souza\IEEEauthorrefmark{2}, \\Nurul Huda Mahmood\IEEEauthorrefmark{1}, Markku Juntti\IEEEauthorrefmark{1}, Matti Latva-Aho\IEEEauthorrefmark{1} }
    \IEEEauthorblockA{\IEEEauthorrefmark{1}Centre for Wireless Communications, University of Oulu, Oulu, Finland 
    \IEEEauthorblockA{\IEEEauthorrefmark{2}Department of Electrical and Electronics Engineering, Federal University of Santa Catarina, Florianópolis, Brazil \\
    \{gabriel.martinsdejesus, onel.alcarazlopez\}@oulu.fi, richard.souza@ufsc.br, \\ \{nurulhuda.mahmood, markku.juntti, matti.latva-aho\}@oulu.fi}}
}

\maketitle

\begin{abstract}
Wireless communications systems must increasingly support a multitude of machine-type communications devices, thus calling for advanced strategies for active user detection (AUD). Recent literature has investigated AUD techniques based on compressed sensing, highlighting the critical role of signal sparsity. This study examines the relationship between frequency diversity and signal sparsity in the AUD problem. Single-antenna users transmit multiple copies of non-orthogonal pilots across multiple frequency channels and the base station independently performs AUD in each channel using the orthogonal matching pursuit algorithm. We note that, although frequency diversity may improve the likelihood of successful reception of the signals, it may also damage the channel sparsity level, leading to important trade-offs. We show that a sparser signal significantly benefits AUD, surpassing the advantages brought by frequency diversity in scenarios with limited temporal resources and/or high numbers of receive antennas. Conversely, with longer pilots and fewer receive antennas, investing in frequency diversity becomes more impactful, resulting in a tenfold AUD performance improvement.
\end{abstract}


\section{Introduction}

\IEEEPARstart{W}{ith} the population of connected wireless devices growing by the day, the design of wireless communications systems is becoming increasingly challenging. Electronic devices communicating with each other without any human interaction, i.e., performing \gls{MTC}, are becoming predominant among all  \cite{Shariatmadari:2015:ICM}. Indeed, the development of the fifth generation of wireless systems (5G) has been driven by the need to support a massive number of heterogeneous \gls{MTC} devices,  e.g., $10^6/\text{{km}}^2$, sharing the same radio resources. They are the so-called \gls{mMTC} scenarios \cite{Chen:2021:IJSAC}, typical for applications such as e.g., industry automation and sensing, and distributed ledgers \cite{Mahmood:2020:arXiv, Pokhrel:2020:Access}.

Generally, in the \gls{mMTC} paradigm, communication is established after the \gls{AUD} phase takes place, which is when the set of active users is detected and identified. When multiple devices, or users, are simultaneously active in a given transmitting opportunity, their signals arrive superimposed to the receiver. This complicates the \gls{AUD} process since the resultant composite signal could erroneously resemble a transmission from a user that is, in fact, not transmitting. Fortunately, users transmit sporadically, with only a small fraction of the total becoming active simultaneously, which can be exploited for accurate \gls{AUD}. Specifically, this sparse user activity encourages the utilization of established \gls{CS} techniques \cite{Knoop:2014:EUSIPCO, Monsees:2015:ACSSC, Choi:2017:COMST}, with the setups designed so that users transmit pilot signals before the data, serving the dual purpose of joint \gls{AUD} and channel estimation. \gls{CS}-based \gls{AUD} has been explored recently and new algorithms have been proposed. For instance, the algorithms proposed in \cite{Cui:2020:ITWC} consider temporal correlation in activation patterns of users, outperforming state-of-the-art alternatives. In \cite{Li:2022:ICL} and \cite{Gao:2022:ICL}, channel estimation and data detection algorithms exchange information with the \gls{AUD} algorithm to increase overall performance by also relying on temporal correlation. The authors in \cite{Yang:2023:IWCL} include a false-alarm detector after the \gls{AUD} phase, guaranteeing better performance at lower \gls{SNR} levels.

A key concept of \gls{CS} is the sparsity level of the signal, denoted by $S\triangleq1/K$, where $K$ refers to the count of active users. The \gls{CS} algorithms can efficiently recover the signals when signal sparsity, time resources, and noise conditions are favorable. In short, for a given number of measurements, the signal is more likely to be recovered when the sparsity level is high (i.e., smaller $K$, with $S\to \infty$ when $K=0$) and/or \gls{SNR} is high. The number of measurements required to guarantee a performance level in \gls{CS} algorithms can be determined based e.g., on the expected sparsity level \cite{Tropp:2007:ITIT}. Although committing to these requirements can be challenging in some applications since both the pilot and data must be transmitted within the coherence interval of the channel, this is not necessarily the case for \gls{mMTC} traffic as packets are usually short.

On the other hand, communication diversity is a well-established method for performance boosting \cite{Pokhrel:2020:Access}, and it can manifest in signal transmission or reception strategies. For instance, it might involve replicating transmissions in the frequency or time domain, or by deploying multiple antennas in the transmitter or receiver. The inclusion of redundant replicas of the same message increases the likelihood of the signal being correctly detected. In the case of transmission diversity, such robustness comes at the cost of increased channel utilization. For instance, with frequency transmission diversity, utilizing multiple copies decreases the sparsity level, resulting in interference, leading to potential trade-offs between diversity and sparsity. These trade-offs have been investigated in terms of throughput, packet collisions, and other physical layer aspects in the literature \cite{Kotaba:2018:IWCL, Boyd:2019:SPAWC}, but this remains an open topic in the case of \gls{CS}-based \gls{AUD}, as indicated in \cite{Lopez:2023:PotI}.  

To identify these trade-offs, herein we consider a scenario where the users transmit their non-orthogonal pilots across several orthogonal frequency channels, and the \gls{BS} implements \gls{AUD} before data decoding. We use the \gls{OMP} algorithm to solve the \gls{CS} problem due to its straightforward implementation and extend it to the multichannel case. Specifically, we propose a method in which the {\gls{BS}} performs {\gls{AUD}} independently in each channel, and the final list of detected users is the union of the lists of all channels. We assess the accuracy of this protocol and the standard single-channel case in several setups. We show that having a sparser signal is significantly more beneficial to {\gls{AUD}} than frequency diversity in most cases where the pilot length is short or multiple antennas are available at the \gls{BS}. On the other hand, as the pilot length becomes longer and multi-antenna receive diversity is unavailable, investing in frequency diversity becomes more relevant, and a tenfold \gls{AUD} performance improvement can be achieved.

The remainder of the paper is organized as follows. We describe our system model in Section \ref{sec:system_model}, and present the \gls{AUD} problem and our proposed method in Section \ref{sec:AUD}. We provide numerical results in Section \ref{sec:numerical_results} discussing the sparsity-diversity trade-offs and conclude the paper in Section \ref{sec:conclusions}.

\section{System Model}\label{sec:system_model}
Consider an \gls{mMTC} deployment with uplink frequency channels $\mathcal{F}$, each denoted by $f \in \mathcal{F}$, with $f = \{1, 2, \dots, F\}$. During a preliminary registering phase, the users connect to the network, engaging in key exchange with the \gls{BS}. Subsequently, the \gls{BS} assigns channel resources and unique pilot sequences to the users. A set $\mathcal{Q}_f$ of $Q \triangleq |\mathcal{Q}_f|$ users operate in channel $f$, but $\mathcal{Q}_f \cap \mathcal{Q}_g = \{\emptyset\}$, $f\neq g$, is not always true if users are allowed to transmit replicas in multiple channels. Users are single-antenna devices while the BS is equipped with $M$ antennas. Time is divided into frame slots comprising of $T$ symbols, and with a duration shorter than the coherence interval. We assume perfect time synchronization, and at each frame slot, $T_p$ symbols are dedicated to joint channel estimation and user activity detection. The remaining symbols are dedicated to data transmission. In this 3-phase system, we only tackle the \gls{AUD} in the present work.

In a grant-free scheme, any user may start a transmission at any time, transmitting its assigned pilot followed by its data. Without loss of generality, we assume all the users have a probability $p$ of becoming active, i.e., starting a transmission. This independent and random transmission pattern may lead to multiple users transmitting simultaneously, resulting in a superposition of signals at the receiver. A way to avoid unrecoverable signals is to assign orthogonal pilots to users. However, $T_p \geq Q$ is required for all pilots to be pair-wise orthogonal, which is generally unfeasible in an \gls{mMTC} scenario. Instead, a list of non-orthogonal pilots is considered. We denote by \mbox{\boldmath$\phi$}$_q\in\mathbb{C}^{T_p \times 1}$ the pilot sequence assigned to the $q$-th user, with $||$\mbox{\boldmath$\phi$}$_q||_2=1$, while the sequences can be arranged to obtain the pilot matrix $\mathbf{\Phi}_f = [$\mbox{\boldmath$\phi$}$_1 \;$\mbox{\boldmath$\phi$}$_2\; \dots\; $\mbox{\boldmath$\phi$}$_{Q}] \in \mathbb{C}^{T_p\times Q}$. 

The channel gains between user $q\in\mathcal{Q}_f$ and the \gls{BS} in channel $f$ are modeled by considering the impact of a large-scale fading term, $\beta_{q,f}$, and a small-scale fading term, $\mathbf{h}_{q,f}\in \mathbb{C}^{1\times M}$. Each element of $\mathbf{h}_{q,f}$ corresponds to the fading realization at each antenna in the \gls{BS}. The expected value of $|\mathbf{h}_{q,j}|^2$ is $\mathbb{E}[|\mathbf{h}_{q,j}|^2]=1$, such that $\mathbf{g}_{q,f} = \beta_{q,f} \mathbf{h}_{q,f}$ \cite{Senel:2018:ITC}.
We assume that $\mathbf{h}_{q,f}$ is unknown and changes at each coherence interval, while $\beta_{q,f}$ is known by the users, and power control can be implemented to result in the desired average \gls{SNR} $\Gamma$ at the \gls{BS}. By arranging the channel gain vectors and taking into account the transmit power employed by users, we obtain the system matrix $\mathbf{X}_f\in \mathbb{C}^{Q\times M}$, with each row given by $\mathbf{x}_{q,f}=\sqrt{T_p\rho_{q,f}}\,\alpha_q\mathbf{g}_{q,f}$, where $\alpha_q=1$ when the $q$-th user is active, $\alpha_q=0$ otherwise, and $\rho_{q,f}$ is the transmit power. Then, at each frame slot, the received signal $\mathbf{Y}_f  \in \mathbb{C}^{T_p\times M}$ at the \gls{BS} in channel $f\in\mathcal{F}$, is given by 
\begin{equation}
    \mathbf{Y}_f = \mathbf{\Phi}_f\mathbf{X}_f + \mathbf{V}_f,
\end{equation}
where $\mathbf{V}_f \in \mathbb{C}^{T_p\times M}$ represents complex white Gaussian noise. Since the noise has unitary variance, the resulting average \gls{SNR} at reception is then controlled by setting $\rho_{q,f}={\Gamma}/{{\beta_{q,f}}^2}$.

\begin{figure}
    \centering
    \includegraphics[]{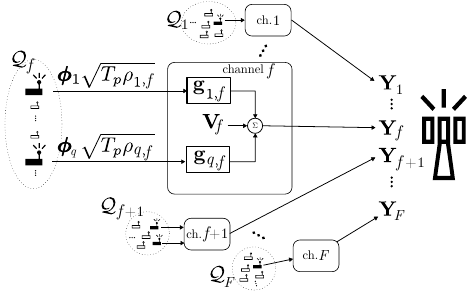}
    \caption{System model with emphasis on channel $f$.}
    \label{fig:system_model}
\end{figure}

We illustrate the system model in Fig. \ref{fig:system_model} explicitly for channel $f$, but the model is valid for the other channels. Note that the grouping of the devices is not representative of their physical location and that our model admits the possibility of a single user operating in multiple different channels.

\section{Active User Detection}\label{sec:AUD}
To effectively decode the received signal, the \gls{BS} must first identify the users that are active in the corresponding frame, for which it can utilize its prior knowledge of the pilot sequences of the users. Specifically, on channel $f$, the \gls{BS} must identify the non-zero rows of $\mathbf{X}_f$. Since the number of rows of $\mathbf{X}_f$ is $Q\gg T_p$, this is an under-determined system, thus without a unique solution in general. However, at each frame slot, only a subset $\mathcal{K}_f\subset\mathcal{Q}_f$ of users are active, making the matrix $\mathbf{X}_f$ to be row-sparse, with  $K=|\mathcal{K}_f|$ denoting its sparsity. This problem can be solved by applying well-established \gls{CS} techniques, such as the \gls{OMP} algorithm. With these techniques, the \gls{BS} can perform joint \gls{AUD} and channel estimation to obtain $\hat{\mathcal{K}}_f$ and  $\hat{\mathbf{X}}_f$, the estimate of the set of active users and an estimate of their channel conditions, respectively.

\subsection{Multi-Channel AUD with Frequency Diversity}\label{sec:proposed_method}
To explore frequency diversity, we propose a protocol in which users transmit copies of the same packets simultaneously across multiple channels. Let each user $q$ be assigned a subset $\mathcal{C}_q\subseteq\mathcal{F}$ of channels for transmission. Each user has access to exactly $C=|\mathcal{C}_q|$ channels, and the list of users operating in each channel is known at the \gls{BS}. To maintain the same spectral efficiency of the single copy case, we consider that the total number of users is $QF$ such that $QC$ users operate in each channel while maintaining the same activation probability $p$. Furthermore, the transmit power per channel is inversely proportional to the number of copies, i.e., ${\rho}_{q,f}' = {{\rho}_{q,f}}/{C}$, so that users adhere to the same power constraints as in the single-frequency method, leading to the target average SNR after reception. 

The copies introduced decrease the sparsity of the received signal by a factor of $C$. On the other hand, they also increase the chances of the signal being received with a favorable fading realization, making it possible to achieve a better \gls{SNR}. We use the discussion from \cite{Wang:2015:ITSP} to make the case for investigating the trade-off between sparsity and diversity in \gls{OMP}-based \gls{AUD}. Let us consider the case with $F=C$, such that $\mathbf{\Phi}_f=\mathbf{\Phi}$ for any $f$, and let us assume that $\mathbf{\Phi}$ is constructed such that it satisfies the restricted isometry property \cite{Candès:2006:CoPAM}. We define ${\Gamma_f}\triangleq{||\mathbf{\Phi}\mathbf{X}_f||^2_2}/{||\mathbf{V}_f||^2_2}$ to be the \gls{SNR} measured at the \gls{BS} in channel $f$. For a signal with sparsity $S=1/K$, a necessary \gls{SNR} condition for perfect signal recovery is given by \cite{Wang:2015:ITSP}
\begin{equation}
    \sqrt{\Gamma_f} \geq \frac{\sqrt{K}(1+\delta_{K+1})}{(1-\sqrt{K}\delta_{K+1})\sqrt{\text{MAR}}},
\end{equation}
where $\delta_{K+1}$ is the isometry constant of $\mathbf{\Phi}$ of degree $(K+1)$ and ${\text{MAR}} \triangleq \min_i ||\mathbf{x}_{i,f}||_2^2/(||\mathbf{X}_f||_2^2/K)$ is the minimum-to-average signal power ratio.

We focus on $\delta_{K+1}$ and note that, the smaller its value, the smaller \gls{SNR} is necessary to recover the active user.  Let $\mathbf{\Phi}^{C_1}\in \mathbb{C}^{T_p\times QC_1} $ and  $\mathbf{\Phi}^{C_2}\in \mathbb{C}^{T_p\times QC_2}$ be random pilot matrices for the $F=C=C_1$ and $F=C=C_2$ cases, respectively, following the same construction rules,  with  $C_2>C_1$. If we consider the same activity probability of devices in both cases, \begin{equation}
    \mathbb{E}[K|C_1]= pQC_1 < \mathbb{E}[K|C_2] = pQC_2.
\end{equation}From the non-decreasing property of the isometry constants \cite{Wang:2015:ITSP}, $\mathbb{E}[\delta_{pQC_1+1}] < \mathbb{E}[\delta_{pQC_2+1}]$, given that they exist. Thus, setups with fewer copies would need a smaller \gls{SNR} for perfect recovery, increasing the likelihood of success.

While having fewer copies decreases the required \gls{SNR} for proper detection, the inclusion of copies can also provide an increase in the \gls{SNR}. For instance, in the same $F=C$ case considered before, one can use the channel with the highest \gls{SNR} to perform the detection, obtaining 
\begin{equation}
    \Gamma_{C} = \max \left\{ \frac{||\mathbf{\Phi}\mathbf{X}_1||^2_2}{||\mathbf{V}_1||^2_2},  \frac{||\mathbf{\Phi}\mathbf{X}_2||^2_2}{||\mathbf{V}_2||^2_2}, \dots, \frac{||\mathbf{\Phi}\mathbf{X}_C||^2_2}{||\mathbf{V}_C||^2_2} \right\}.
\end{equation}
Then, for $C_2>C_1$, we have $\mathbb{E}[\Gamma_{C_2}]\geq \mathbb{E}[\Gamma_{C_1}]$, as there are more realizations to choose from, thus making it more likely that $\Gamma_C$ has a higher value. This gain in \gls{SNR}, contrasted with the loss in sparsity, suggests a trade-off between the two. While this discussion motivates our search for this trade-off, the isometry constant is known to be hard to obtain, and there are other possible ways of exploring the frequency diversity. We proceed our analysis with computer simulations as we consider more complex combinations of $F$ and $C$.

Several approaches can be implemented to explore the multiple-channel setup. We have implemented four such approaches, three of which did not show any improvement when compared to the single-channel case. We list them below and briefly comment on their possible advantages and the cause of their poor performance when that is the case.
\begin{enumerate}
    \item Strict detection:  perform the \gls{AUD} in each channel independently, but only accept users detected in all their corresponding channels. This results in a low number of false positives but also limits the performance by not considering users that had their signals too attenuated in other channels.
    \item Iterative detection: perform the \gls{AUD} sequentially in each channel, detecting users and propagating their detection to the other channels where they transmit, iteratively removing their contribution to the received signal. This requires fewer operations at the \gls{BS}, as each user needs to be detected only once, but causes many false positives. 
    \item Super-channels: combine the channels into ``super-channels'' by concatenating the received signals and extending the pilot matrix accordingly. This leads to $\binom{F}{C}$ such channels, and the \gls{BS} searches only for the users operating exclusively in the $C$ channels considered, treating signals of other users as interference. This increases the number of measurements available, but, when $C\neq F$, the interference from other users causes poor performance. When $C=F$, the extended pilots are not helpful in further differentiating users, causing no improvements.
    \item\label{app:independent} Independent detection: perform the \gls{AUD} in each channel independently, constructing a final list consisting of the union of all lists. With such an implementation, one can configure the algorithm to detect few users in each channel, which causes fewer false positives, as only the most likely users in each channel are selected. The result is a better balance of missed detections and false positives, while being much simpler than the others in its implementation. 
\end{enumerate}
In Section \ref{sec:numerical_results}, we compare these approaches in Fig. \ref{fig:comparison} for the configuration $F=C=2$. For brevity, in this paper, we focus on approach \ref{app:independent}, which performs better than the others. As we will show in the sequence, this approach can achieve better results than the conventional method in some scenarios.

\subsection{Orthogonal Matching Pursuit}
Originally proposed in \cite{Pati:1993:ACSSC}, the \gls{OMP} is a greedy algorithm for \gls{CS}. Due to its simple implementation and manageable computational complexity, \gls{OMP} serves as a clear baseline for other algorithms, and thus we adopt it to evaluate our proposal. The basic idea behind \gls{OMP} is to select the columns of pilot matrix $\mathbf{\Phi}_f$ that have the highest correlations with the received signal $\mathbf{Y}_f$ and iteratively remove their contribution until a stopping criterion is satisfied.

The \gls{OMP} receives as inputs the signal  $\mathbf{Y}_f$ and the pilot matrix $\mathbf{\Phi}_f$. 
Initially, the list of detected users is $\hat{\mathcal{K}}_f=\emptyset$, and the residual signal is $\mathbf{R}_f^{(0)} = \mathbf{Y}_f$. At each iteration $k$, the correlation $r_{q,f}^{(k)} \in \mathbb{C}^{1 \times M}$ of the residual signal to the pilot sequence of user $q$ is calculated as
\begin{equation}
    r_{q,f}^{(k)} = ||\mbox{\boldmath$\phi$}_q^\intercal \mathbf{R}_f^{(k-1)}||_2.
\end{equation}
Then, the \gls{BS} identifies the user whose pilot sequence has the highest correlation with the residual signal as 
\begin{equation}
    q^{(k)} = \underset{q}\argmax\, r_{q,f}^{(k)},
\end{equation}
updating the  list of detected users as $\hat{\mathcal{K}}^{(k)}_f = \hat{\mathcal{K}}^{(k-1)}_f \cup q^{(k)}.$ At each iteration $k>1$, an auxiliary pilot matrix $\mathbf{\Phi}_{f}^{(k)}\in\mathbb{C}^{T_p \times k}$ is updated as
\begin{equation}\label{OMP:update_phi}
    \mathbf{\Phi}_{f}^{(k)} =  \left[\mathbf{\Phi}_{f}^{(k-1)},\, {\boldsymbol \phi}_{q^{(k)}} \right],
\end{equation}
with $\mathbf{\Phi}_{f}^{(1)} \triangleq {\boldsymbol \phi}_{q^{(1)}}$. Then, an estimate of the channel is obtained by solving
\begin{equation}\label{OMP:update_signal}
    \hat{\mathbf{X}}^{(k)}_{f} = \underset{\mathbf{X}}\argmin ||\mathbf{Y}_f - \mathbf{\Phi}_{f}^{(k)}{\mathbf{X}}||_2.
\end{equation}
In our implementation, we solve it with the linear least squares method, thus the channel is estimated with
\begin{equation}
    \hat{\mathbf{X}}^{(k)}_{f} = {\mathbf{\Phi}_{f}^{(k)}}^+\mathbf{Y}_f,
\end{equation}
where ${(\cdot)}^+$ is the Moore-Penrose inverse operation. The residual signal is then updated with
\begin{equation}\label{OMP:update_residual}
    \mathbf{R}_{f}^{(k)} = \mathbf{Y}_{f} - \mathbf{\Phi}_{f}^{(k)}\hat{\mathbf{X}}^{(k)}_{f},
\end{equation}
finalizing the $k$-th iteration.

The \gls{OMP} continues to run until the stopping criterion is met, and the final list of detected users in channel $f$ is denoted by $\hat{\mathcal{K}}_f$. When $K$ is known or estimated at the \gls{BS} \cite{Lopez:2023:ITC}, the stopping criterion is $k=K$. When this value is unknown, a different approach should be considered. We consider a residual-based approach \cite{Choi:2017:COMST}, allowing the algorithm to run while $r_{q,f}^{(k)}\geq\varepsilon$ for any $q$. We run training simulations where the $K$ users are known to obtain a range of viable values for $\varepsilon$. Then, for a fair comparison between each configuration and algorithm considered, $\varepsilon$ is selected individually to achieve maximum accuracy in each case. In practice, $\varepsilon$ must be carefully selected, as large values lead to many false negatives, while small values lead to many false positives, compromising the accuracy.

\section{Numerical results}\label{sec:numerical_results}
In this section, we analyze the performance of the considered system when varying three main network parameters, namely the number of copies $C$ each user transmits, the length $T_p$ of the pilot sequence, and the activation probability $p$. 
We define the balanced accuracy $A$ as 
\begin{equation}
    A = \frac{1}{2}\left(\frac{|\cup_{f\in\mathcal{F}} \text{TP}_f|}{K} + \frac{|\cup_{f\in\mathcal{F}} \text{TN}_f|}{Q-K}\right),
\end{equation}
where $\text{TP}_f = \hat{\mathcal{K}}_f \cap \mathcal{K}_f$ is the list of active users correctly identified, and  $\text{TN}_f = (\mathcal{Q}_f\backslash\hat{\mathcal{K}}_f) \cap (\mathcal{Q}_f\backslash\mathcal{K}_f)$ is the list of inactive users correctly identified. The balanced accuracy is particularly well-suited for classifiers dealing with unbalanced class distributions, which is the case in \gls{mMTC} since $K\ll (Q-K)$. We consider its complement, the balanced inaccuracy $1-A$, as the main performance metric. We present the average balanced inaccuracy for $10^5$ Monte-Carlo simulations for each case. The final results are the minimum achieved in each setup, with the threshold $\varepsilon$ optimized for each configuration. 

The pilots are generated by drawing symbols from a complex Bernoulli distribution \cite{Senel:2018:ITC}, with each sequence designed such that $||$\mbox{\boldmath$\phi$}$_q||_2=1$ and each entry is randomly selected from the possible symbols $(\pm 1 \pm i)/\sqrt{2T_p}$. Assuming the channels are subject to Rayleigh fading, each element of $\mathbf{h}_{q,f}$ is distributed following a standard complex normal distribution. Moreover, we set $Q=256$ users and $p=1/Q$, unless specified otherwise. 

\begin{figure}
    \centering
    \includegraphics[width=0.5\textwidth]{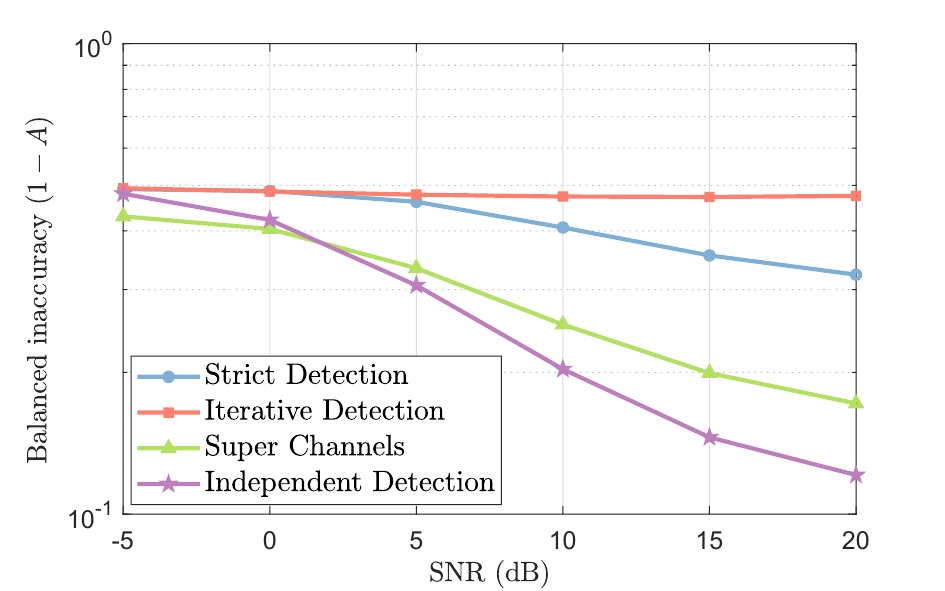}
    \caption{A comparison of each of the approaches listed in Section \ref{sec:proposed_method}.}
    \label{fig:comparison}
\end{figure}

In Figure \ref{fig:comparison}, we present a comparison of the methods outlined in Section \ref{sec:proposed_method}, for the configuration with $F=C=2$. While these results may not be exceptional, the Independent Detection method still outperforms the others considered, warranting further investigation. Herein, we compare in details this method to the classical approach.

\subsection{Impact of the Number of Copies}
\begin{figure}
    \centering
    \includegraphics[width=0.5\textwidth]{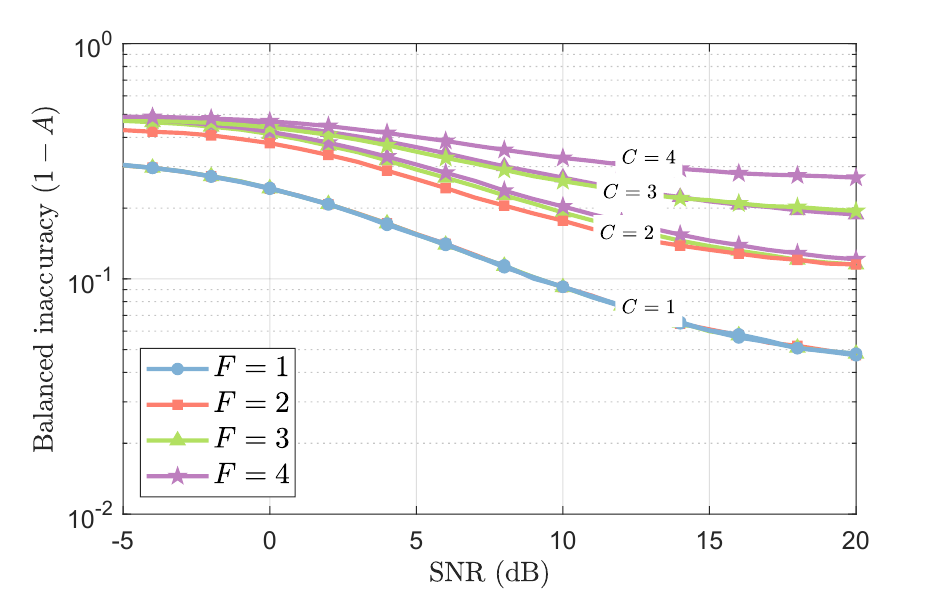}
    \caption{Balanced inaccuracy as a function of the SNR and for several combinations of $F$ and $C$ in the $M=1$ setup.}
    \label{fig:number_of_copies_M1}
\end{figure}

We start by evaluating the potential performance gains from diversity in the base scenario described earlier. The number of channels is varied from $F=1$ to $F=4$, with the corresponding number of copies varied from $C=1$ to $C=F$, for several values of \gls{SNR} and for $M=1$ and $T_p=8$. The results presented in Fig. \ref{fig:number_of_copies_M1} indicate that, in this specific setup, the added diversity does not improve the system performance but worsens it as the number of copies increases. Here, the sparsity level decreases inversely proportional to $C$, and the pilot length is not sufficient to provide significant differentiation among the users, causing the \gls{BS} to often mistakenly select inactive users, as their pilot may be correlated to the superimposed signal. An interesting behavior in this case is the similar performance regardless of $F$, but dependent on $C$, as this parameter is what impacts the sparsity level in each channel. This behavior can be further observed for $F>4$, but we use $F=4$ as the default herein.

\subsection{Impact of the Pilot Length}
\begin{figure}
    \centering
    \includegraphics[width=0.5\textwidth]{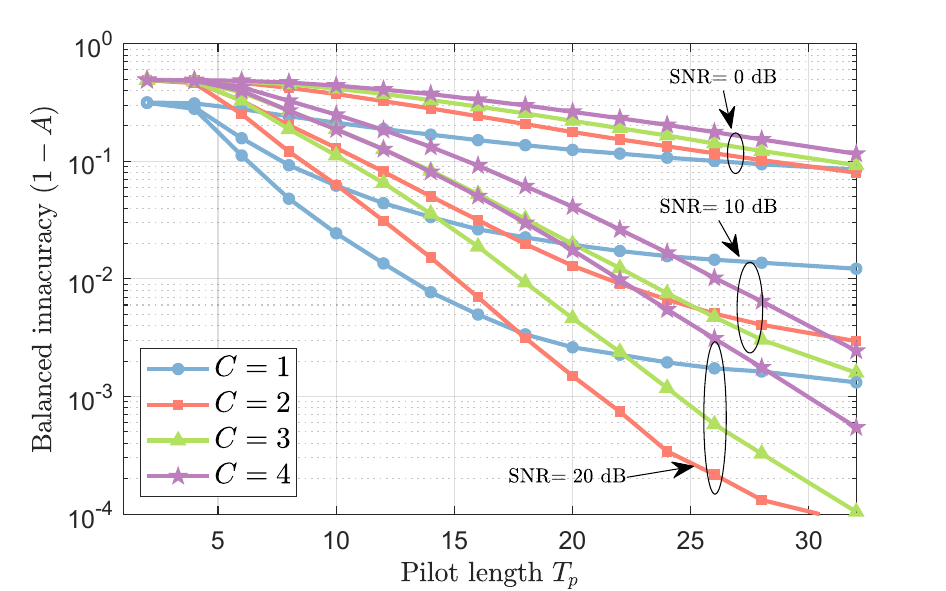}
    \caption{Balanced inaccuracy as a function of the pilot length in the $M=1$ setup.}
    \label{fig:pilot_length_M1}
\end{figure}

We continue our analysis to understand the impact of the pilot length in this new scenario. Since the increase of measurements given by a greater $T_p$ serves the purpose of differentiating more precisely between possibly active users, we vary the pilot length from $T_p=8$ to $T_p=32$. In this new setup, we compare the performance of the system under three \gls{SNR} conditions, for deployments with $F=4$ and $C\leq 4$. For the case with a single antenna \gls{BS} shown in Fig. \ref{fig:pilot_length_M1}, as the pilot length increases, increasing the number of copies is favorable compared to $C=1$, with performance gains of up to one order of magnitude when $\text{{SNR}}=20$~dB. An indiscriminate increase in $C$ is not encouraged, though, as optimal values of $C$ generally vary between $C=2$ and $C=3$. As $T_p$ increases, the frequency diversity gains, that is, the possibility of a copy of a signal to be detected in different channels, are beneficial to the system as it is more unlikely that a false detection takes place. In the meantime, strong signal attenuation in a different channel is not a game-breaker, as it can facilitate other users to be correctly detected. 

\begin{figure}
    \centering
    \includegraphics[width=0.5\textwidth]{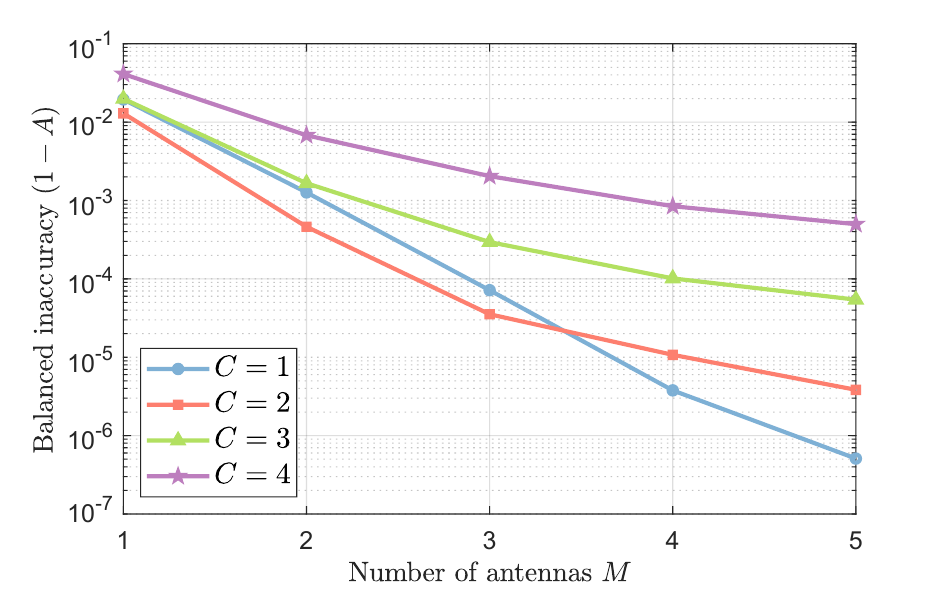}
    \caption{Balanced inaccuracy for number of copies as a function of $M$, in the $T_p=20$, $\text{{SNR}}=10$~dB and $F=4$ setup.}
    \label{fig:pilot_length_M16}
\end{figure}

As indicated in Fig. \ref{fig:pilot_length_M16}, with the increase of $M$, severe signal attenuation is less relevant due to the multiple copies collected, and sparsity is favored, as the more accentuated superposition of the pilots generates more errors when $C>1$. Fundamentally, spatial diversity in reception plays a similar role to frequency diversity in transmission: obtaining redundant replicas of signals to improve user detection likelihood. But, unlike transmission diversity, reception diversity does not incur the drawback of damaging the signal sparsity. This significantly impacts the \gls{AUD}'s performance, discouraging the use of frequency diversity when the \gls{BS} employs multi-antenna reception capabilities. When that is the case, an antenna count of $M=3$ with $T_p=20$ already results in a performance comparable to the setup with $M=1$, $C=2$, and $T_p=32$. In some cases, this is preferred as the signal processing is less demanding and the frame can be used to transmit more data, though it costs more physical resources.

\subsection{Impact of the sparsity level}
\begin{figure}
    \centering
    \includegraphics[width=0.5\textwidth]{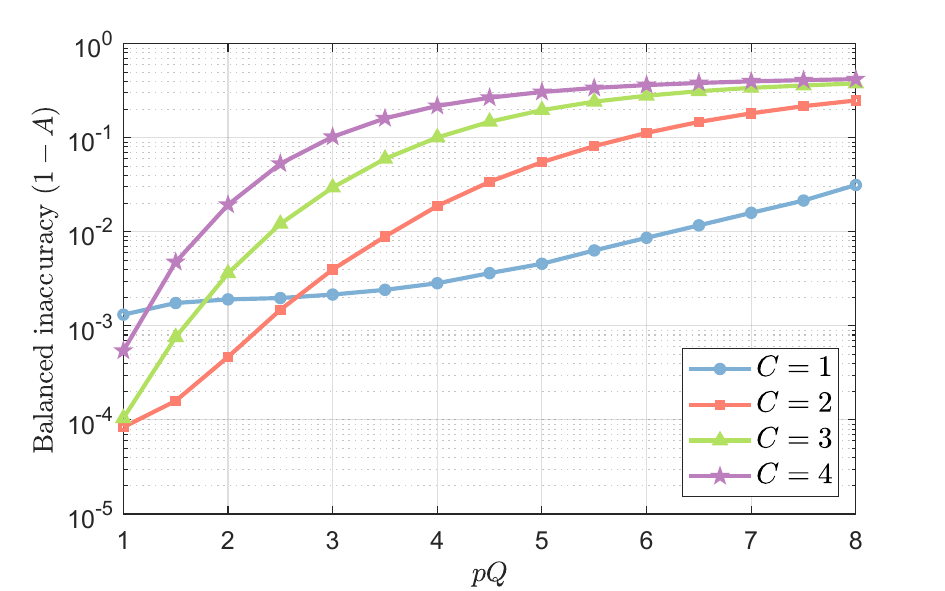}
    \caption{Balanced inaccuracy as a function of  $pQ$ with $M=1$ setup with $\text{{SNR}}=20$~dB and $T_p=32$.}
    \label{fig:sparsity_level_M1}
\end{figure}

Lastly, we vary the probability of activation from $p=0.25/Q$ to $p=8/Q$ in a setup with $F=4$, and $C=\{1, 2, 3, 4\}$, as presented in Fig. \ref{fig:sparsity_level_M1}, where we set $M=1$, $T_p=32$ and $\text{SNR}=20$~dB. The figure can be split into three different regions: $p\leq 1/Q$, $1/Q<p\leq 2.5/Q$, and $p>2.5/Q$. In the first region, the expected value of $K$ is generally below 1,  and the diversity gain introduced by the copies significantly improves performance, especially for higher values of $C$. In the second region, $1<C<4$ has the best performance, with $C=2$ being generally the best option, as with $C=3$ the sparsity levels decline quickly. In the third region, however, the sparsity degradation from the introduced diversity starts to severely limit the performance in all cases where $C\neq 1$, and $C=1$ is once again preferred.

\section{Conclusions}\label{sec:conclusions}
In this paper, we explored the effects of frequency diversity in \gls{CS}-based \gls{AUD}, in a protocol in which users transmit copies of their message across $C$ channels. We maintain the same spectral efficiency when compared to $C=1$, which decreases the sparsity level in each channel. We show that the compromise between frequency diversity and sparsity level hugely impacts the performance of the networks when time resources like pilot length are limited and users transmit frequently, or when there is an increase in the number of receive antennas, and $C=1$ emerges as the preferred choice. However, our results demonstrate the potential of the multi-copy scheme given additional resources allocated to extend the pilot length. For higher \gls{SNR}, adopting this strategy can lead to a performance enhancement of up to one order of magnitude.

\section{Acknowledgement}
This research has been supported by the Finnish Foundation for Technology Promotion and the Research Council of Finland (former Academy of Finland) Grant 346208 (6G Flagship Programme), CNPq (401730/2022-0, 402378/2021-0, 305021/2021-4) and RNP/MCTIC 6G Mobile Communications Systems (01245.010604/2020-14).

\bibliographystyle{IEEEtran}
\bibliography{main}

\end{document}